Spin Polarization and Electronic Structure of Room-Curie Temperature Ferromagnetic $Mn_5Ge_3$ Epilayers

R. P. Panguluri

*Department of Physics and Astronomy, Wayne State University, Detroit, MI 48201*

Changgan Zeng and H.H. Weitering

*Department of Physics and Astronomy, The University of Tennessee, Knoxville, TN 37996*

*and Condensed Matter Sciences Division, Oak Ridge National Laboratory, Oak Ridge, TN 37831*

J.M. Sullivan and S.C. Erwin

*Center for Computational Materials Science, Naval Research Laboratory, Washington, DC 20375*

B. Nadgorny

*Department of Physics and Astronomy, Wayne State University, Detroit, MI 48201*

**Abstract:**

Germanium-based alloys hold great promise for future spintronics applications, due to their potential for integration with conventional Si-based electronics. $Mn_5Ge_3$ exhibits strong ferromagnetism up to the Curie temperature $T_c \sim 295K$. We use Point Contact Andreev Reflection (PCAR) spectroscopy to measure the spin polarization of $Mn_5Ge_3$ epilayers grown by solid phase epitaxy on Ge(111). In addition, we calculate the spin polarization of bulk $Mn_5Ge_3$ in the diffusive and ballistic regimes using density-functional theory. The measured spin polarization, $P_c = 43\pm5\%$, is compared to our theoretical estimates, $P_{DFT} = 10\pm5\%$ and $35\pm5\%$ in the ballistic and diffusive limits respectively.

Semiconductor spintronics has recently been recognized for its potential of combining spin-selectivity and non-volatility of ferromagnetic compounds with the ability to fabricate hybrid thin-film devices naturally integrated with conventional semiconductors.T[1] Several novel applications based on these compounds, such as quantum computing, Datta-Das type spin transistors, and non-volatile memory, can be envisioned[2]. Yet, in order to succeed in technological applications one has to be able to introduce *room-temperature* spin-polarized materials, that are compatible with, and preferably lattice-matched to mainstream GaAs- and/or Si-based electronic materials. $Ga_{1-x}Mn_xAs$ epilayers, fabricated by the Molecular Beam Epitaxy (MBE) technique, satisfy some of these requirements. However, in spite of considerable progress in increasing the Curie temperature of this alloy[3], it has yet to exceed 200 K. Another potential problem for $Ga_{1-x}Mn_xAs$ is the intrinsic disorder of Mn on the Ga sub-lattice due to its non-stoichiometric composition, which limits the hole mobility and consequently spin transport in these compounds. Some reports on GaMnN and other dilute magnetic semiconductors with smaller lattice constants and larger effective valence-band masses suggest the possibility of higher Curie temperatures; however, transport properties in these materials are likely to suffer even more than in GaMnAs.

Recently, *Zeng at. al[4]*. reported the first measurements of a novel ferromagnetic $Mn_5Ge_3$ epitaxial film grown on Ge(111) substrate with Curie temperature $T_c = 296$ K. The film has sufficiently good crystalline quality, surface topography, and thermal stability to allow the growth of germanium-based heterostructures. Although one would ideally like to have Curie temperatures significantly exceeding the operating temperature of a magnetic device, this is an important first step towards this goal. In addition to having high Curie temperature, sufficiently high spin polarization is essential for $Mn_5Ge_3$ to be useful in a spintronics device.

In this Letter we report measurements of the spin polarization for $Mn_5Ge_3$ epilayers using Point Contact Andreev Reflection Spectroscopy (PCAR), and calculations of spin polarization for bulk $Mn_5Ge_3$ in the ballistic and diffusive regimes using density-functional theory (DFT).

Spin polarization in the ballistic and diffusive limits is given by the expression

$$P_n = \frac{<N_\uparrow(E_f)v^n_{f\uparrow}> - <N_\downarrow(E_f)v^n_{f\downarrow}>}{<N_\uparrow(E_f)v^n_{f\uparrow}> + <N_\downarrow(E_f)v^n_{f\downarrow}>},$$

with $n$ = 1 and 2, respectively. Here, $N_\uparrow(E_f)$ and $N_\downarrow(E_f)$, $v_{f\uparrow}$ and $v_{f\downarrow}$ are the DOS and the Fermi velocities for majority and minority spin sub-bands. $P_{1,2}$ can be measured by the Andreev reflection technique[5,6], particularly in the point contact configuration (PCAR) recently introduced to measure the spin polarization of ferromagnets[7]. The PCAR technique is based on the difference in the number of channels available for Andreev reflection process[8] in a normal metal/ superconductor (N/S) junction[8] and in the ferromagnet/superconductor (F/S) junction[9], making it possible to infer the spin polarization of a ferromagnet by analyzing the conductance curves using the appropriate weak coupling theory.[10]

In spite of a reasonable success of the PCAR techniques in magnetic metals, there are some theoretical difficulties concerning rigorous description of the Andreev process, as was pointed out by Xia *et al.*[11] While recent works on Andreev reflection in semiconductors consistently indicate conventional behavior in the case of non-magnetic alloys, such as $Ga_{1-x}Be_xAs$ and $In_{1-y}Be_ySb$[12,13], as well as magnetic $In_{1-z}Mn_zSb$, the first studies of $Ga_{1-x}Mn_xAs$ epilayers,[12,14] seem to indicate the presence of some unconventional processes, near the semiconductor/superconductor interface. Comparison between $Ga_{1-x}Be_xAs$, $Ga_{1-y}Mn_ySb$, and $In_{1-z}Mn_zSb$[15] points to the presence of inelastic scattering processes as a possible culprit. These processes, in turn, become more pronounced with increased strength of magnetic fluctuations and decreased mobility in magnetic semiconductors and are dependent upon the degree of disorder in dilute magnetic semiconductors. From this perspective, the $Mn_5Ge_3$ system has an important advantage: It is a stoichiometric material, so its Mn sub-lattice is well ordered and has higher mobility. The fact that $Mn_5Ge_3$ is a stoichiometric compound also makes it possible to use DFT to determine the densities of states, Fermi velocities, and spin polarization — thereby allowing a direct comparison of theoretical and experimental results.

$Mn_5Ge_3$ films with a typical thickness of 50 nm were grown by depositing Mn onto Ge(111) using an (MBE) process with a base pressure of $4.0 \times 10^{-11}$ mbar, and by subsequent annealing for several minutes, a technique known as solid phase epitaxy (SPE). $Mn_5Ge_3$ films were always uniform when annealed between 300°C and 650°C. X-ray diffraction $\theta$-$2\theta$ scan results indicate good quality epitaxial film. The intermetallic compound $Mn_5Ge_3$ has the hexagonal crystal structure with the unit cell parameters at room temperature $a = 7.184$ Å and $c = 5.053$ Å. The stoichiometry of 0.4 for Mn obtained from Rutherford Backscattering measurements is close to 0.375, corresponding to the $Mn_5Ge_3$ structure, possibly with some Mn deficiency. Further details of the growth procedure can be found in Ref. 4.

The temperature dependence of the magnetization measured by a SQUID magnetometer demonstrates a fairly conventional magnetic behavior shown in Fig. 1 (the easy axis of the film is in plane due to shape anisotropy). The Curie temperature is ~ 295 K, which is practically the same as that of bulk $Mn_5Ge_3$. The temperature dependence of the resistivity $\rho$ is metallic with $\rho_{4K}$ ~ 10 μΩ.cm.

Mechanically sharpened Sn superconductor tips were used for all point contact measurements in this study. A contact was established between the sample and the Sn tip along the [001] direction of $Mn_5Ge_3$. The conductance was measured by the standard four-terminal technique with a lock-in detection at 2kHz. Conductance curves were fitted using the modified BTK model [10] with two fitting parameters, the spin polarization, $P_c$ and the dimensionless interfacial scattering parameter, $Z$. The details of the measurement technique are given in Ref.16. In contrast to the $Ga_{1-x}Mn_xAs$ measurements, a series of measurements in all of the $Sn/Mn_5Ge_3$ point contacts in several different samples demonstrate a conventional behavior, with the superconducting gap (used in the data analysis) close to the bulk value in Sn, $\Delta(0)$ ~ 0.55mV. A representative $dI/dV$ curves for two different contacts are shown in Fig. 2 (a,b). The gap at higher temperatures was consistently obtained from the BCS $\Delta(T)$ dependence. The spin polarization for $Mn_5Ge_3$, obtained by averaging over many different contacts and temperatures, was found to be $P_c$ ~ 43±5%. Fairly large uncertainty in the measured spin polarization is likely to arise from strong crystallographic anisotropy, resulting in strong directional dependence of $P_c$.[17]

The theoretical calculations were carried out within the generalized-gradient approximation (GGA) to DFT, using the projector-augmented wave (PAW) method and a plane-wave basis.[18] All the calculations used an energy cutoff of 337 eV; the structural optimization and DOS calculations used 4×4×6 and 12×12×14 Γ-centered samplings, respectively, of the $Mn_5Ge_3$ hexagonal Brillouin zone. The resulting equilibrium lattice parameters, $a=7.092$ Å and $c=4.984$ Å, are in good agreement with our experimental results, and the two equilibrium internal parameters, $x(Mn)=0.244$ and $x(Ge)=0.606$, are very close to experimental values reported earlier for $Mn_5Ge_3$, $x(Mn)=0.2397$ and $x(Ge)=0.6030$.[17] The resulting $Mn_5Ge_3$ band structure is shown in Fig. 3a. Based on this band structure, the spin polarization was calculated according to the definitions given earlier, using the linear tetrahedron method. For the $z$-direction (perpendicular to the hexagonal plane) we obtain $P_1 = -9\%$ and $P_2 = +35\%$ (see Fig. 3b). We estimate the overall numerical uncertainty in these values to be on the order of 5%. Interestingly, when we repeat the calculations using the experimental lattice and internal parameters (which differ by 1-2% from the theoretical values) we obtain very different spin polarizations, $P_1 = +15\%$ and $P_2 = +50\%$, indicating a surprisingly large sensitivity of the polarization to the details of the crystal structure. Crystallographic anisotropy may also be important: the in-plane ballistic spin polarization, $P_{1\parallel} = +10\%$, is very different from that along the $z$-direction, $P_{1\perp} = -9\%$. The fact that $P_1$ changes sign between the in-plane and $z$-directions indicates that at some intermediate angle the spin polarization should be close to zero.[18]

Using $\sigma_i = e^2 \langle N(E_F) v_{Fi}^2 \rangle \tau$ and the known conductance of the $Mn_5Ge_3$ film we can estimate the mean free path $L$. From the calculated values of $[\langle N(E_F) v_{Fz\uparrow}^2 \rangle + \langle N(E_F) v_{Fz\downarrow}^2 \rangle] = 8.1$ eV·Å$^2$ and $<v_{Fz\uparrow}> \sim 1.7 \cdot 10^7$ cm/s, $<v_{Fz\downarrow}> = 2.0 \cdot 10^7$ cm/s we can find the relaxation time $\tau = L/<v_F> \sim 4.5 \cdot 10^{-13}$ s, yielding $L \sim 85$ nm. The contact size contact $d$ can be estimated from the Sharvin formula $R_n \sim 4\rho L / 3\pi d^2 + \rho/2d$, where $R_n$ is the contact resistance. For the typical values of the contact resistance $R_n \sim 30 \Omega$ we have obtained the contact size $d \sim 25$ nm, indicating that our measurements were done in the ballistic regime, $L >> d$.

In summary, we have measured the transport spin polarization of the *room Curie temperature* epitaxial $Mn_5Ge_3$ film. All the measurements were done in the ballistic regime. The measurements indicate fairly conventional Andreev reflection in high transparency junctions, with small interfacial scattering parameter, $Z$. The spin polarization of $Mn_5Ge_3$ along the [001] direction was measured with PCAR to be 43±5%, much higher than the $P_{DFT}$ = 10-15% in the ballistic limit, but fairly close to the $P_{DFT}$ = 35-50% in the diffusive limit. The agreement with the latter results must be coincidental as all the measurements are done in the ballistic regime. We attribute the discrepancy between the experiment and the theory to extreme sensitivity of the DFT calculations to the crystallographic structure of $Mn_5Ge_3$, as well as possible Mn deficiency of $Mn_5Ge_3$. Overall, our results demonstrate that not only $Mn_5Ge_3$ can be lattice-matched to Ge, has high mobility and room Curie temperature, but it also has higher than predicted spin polarization, comparable to the spin polarization of 3-d metals as well as MnAs, indicating high potential for this novel material in a variety of spintronics applications.

We thank Y. Lyanda-Geller, I. Mazin, and I. Zutic for useful discussions and S. Picozzi for sharing the results of their calculations in $Mn_5Ge_3$ prior to publication. Computations were performed at the DOD Major Shared Resource Center at ASC. The work is supported by DARPA through ONR Grant N00014-02-1-0886 and NSF Career Grant 0239058 (B.N.), and by ONR (S.E.). H.H.W. is supported by NSF under DMR Grant 0306239 (FRG). Oak Ridge National Laboratory is managed by UT-Battelle, LLC, for the U.S. DOE under Contract No. DE-AC05-00OR22725.

Figure Captions.

Fig. 1. Temperature dependence of the in-plane magnetization for the $Mn_5Ge_3$ epitaxial film with the Curie temperature ~ 295K. Magnetic field is also in-plane. The cusp at ~67K indicate another magnetic transition at this temperature (see Ref. 4).

Fig. 2. Typical normalized conductance curves Sn superconducting contacts with $Mn_5Ge_3$ epitaxial film shown for the minimum (a) and maximum (b) measured spin polarization: (a) Contact resistance $R_c$ = 32Ω, $T$ = 1.2K, $\Delta$ = 0.55mV. Fitting parameters: $Z$ = 0.2, $P$ = 33%; (b) Contact resistance $R_c$ = 8Ω, $T$ = 1.6K, $\Delta$ = 0.5 mV. Fitting parameters: $Z$ = 0.1, $P$ = 46%.

Fig.3. a) Theoretical band structure of $Mn_5Ge_3$ for the optimized case. Open circles and solid curves denote majority bands; filled circles and dotted curves denote minority bands. B) Spin Polarization for the DOS ($P_0$), ballistic case ($P_1$), and diffusive case ($P_2$).

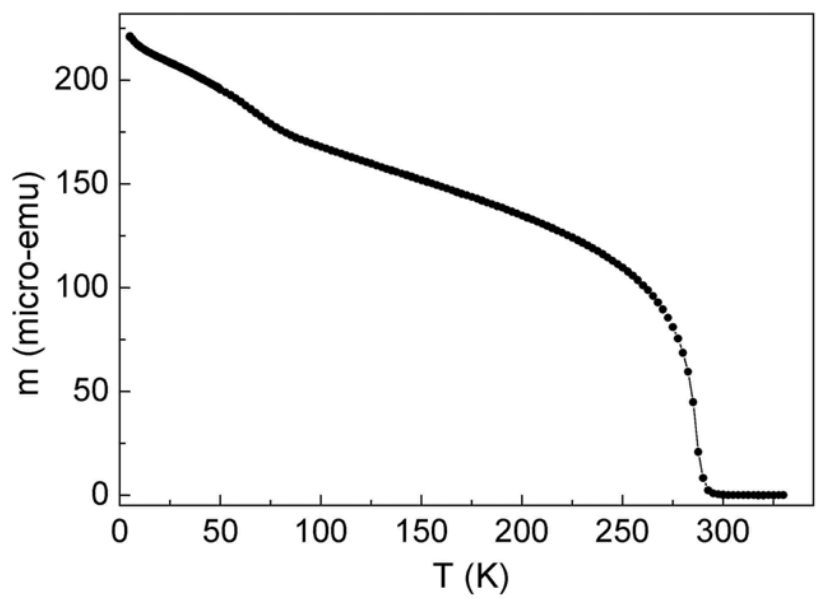

**Figure 1. Panguluri et al.**

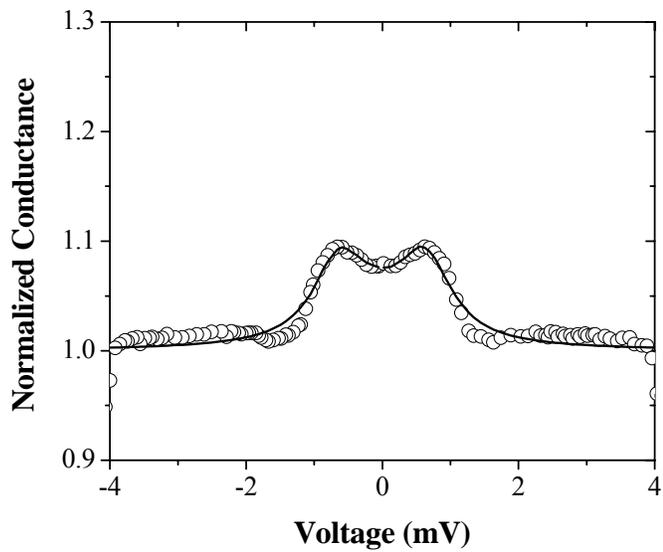 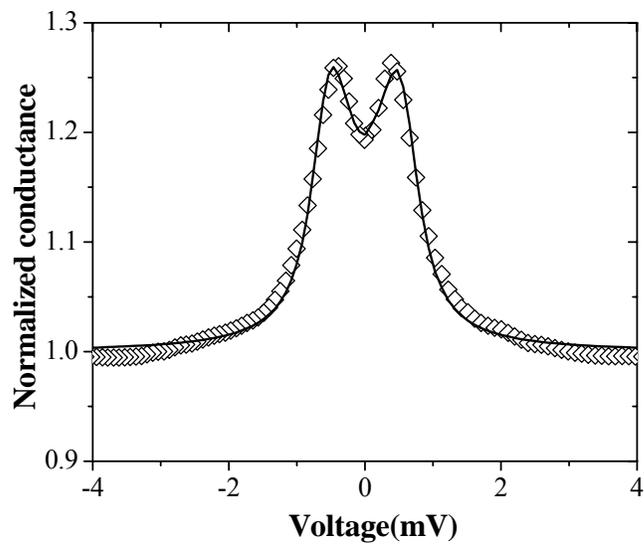

**(a)** **(b)**

**Figure 2 (a,b).** **Panguluri et al.**

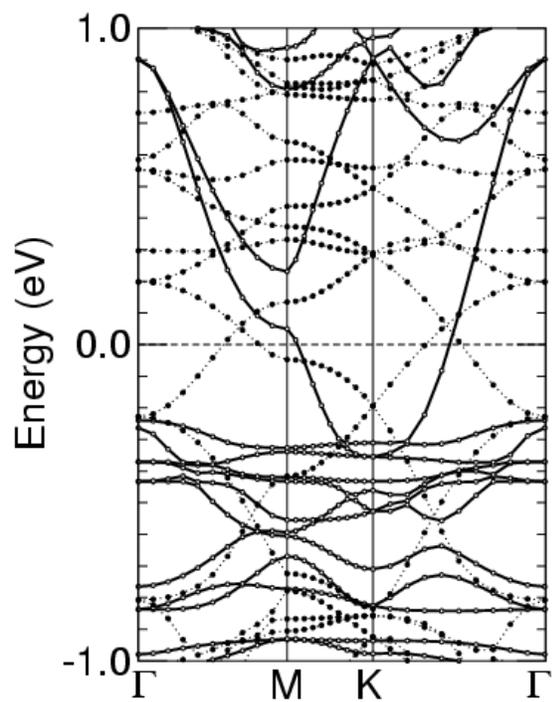 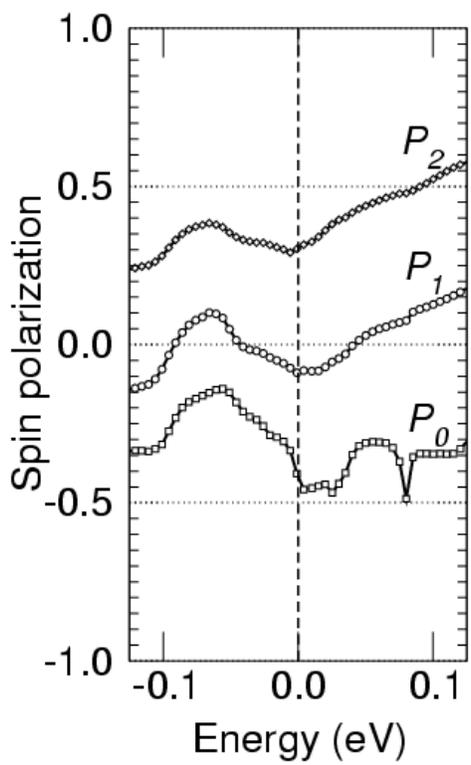

(a)          (b)

Figure 3 (a,b).          Panguluri et al.

REFERENCES:

[1] H. Munekata, H. Ohno, S. von Molnar, A. Segmüller, L. L. Chang, and L. Esaki, *Phys. Rev. Lett*. **63**, 1849 (1989).

[2] S.A. Wolf, D.D. Awschalom, R.A. Buhrman, J.M. Daughton, S. von Molnar, M.L. Roukes, A.Y. Chtchelkanova and D.M. Treger, *Science* **294**, 1488 (2001).

[3] A.M. Nazmul, S. Sugahara, M. Tanaka, *J. Cryst. Growth* **251,** 303 (2003).

[4] C. Zeng, S. C. Erwin, L. C. Feldman, A.P. Li, Y. Song, J.R. Thompson, and H. H. Weitering, *Appl. Phys. Lett*. **83**, 5002 (2003).

[5] R.J. Soulen, Jr., J.M. Byers, M.S. Osofsky, B. Nadgorny, T. Ambrose, S.F. Cheng, P.R. Broussard, C.T. Tanaka, J. Nowak, J. S. Moodera, A. Barry, and J.M.D. Coey, *Science*, **282**, 85-88 (1998).

[6] S.K. Upadhyay, A. Palanisami, R.N. Louie, and R.A. Buhrman, *Phys. Rev. Lett*. **81**, 3247-3250 (1998).

[7] I. Zutic, J. Fabian, and S. Das Sarma, *Rev. Mod Phys*., **76**, 323 (2004).

[8] A.F. Andreev, *Sov. Phys. JETP* **19**, 1228 (1964).

[9] M.J.M. de Jong and C.W.J. Beenakker, *Phys. Rev. Lett*. **74**, 1657-1660 (1995).

[10] I.I. Mazin, AA. Golubov and B. Nadgorny, *Journ. Appl. Phys*. **89**, 7576 (2001).

[11] K.Xia, P.J. Kelly, G.E.W. Bauer, and I. Turek, *Phys. Rev. Lett.,* 89, 166603 (2002).

[12] R.P. Panguluri, K.C. Ku, N. Samarth, T. Wojtowicz, X. Liu, J.K. Furdyna, Y. Lyanda-Geller, and B. Nadgorny, *submitted to PRL*.

[13] R. P. Panguluri, B. Nadgorny, T. Wojtowicz, W.L. Lim, X. Liu, and J.K. Furdyna, *Appl. Phys. Lett.,* **84**, 4947 (2004).

[14] J.G. Braden, J.S. Parker, P. Xiong, S.H. Chun, and N. Samarth, *Phys. Rev. Lett.* **91**, 056602 (2003).

[15] T. Wojtowicz, G. Cywinski, W. L. Lim, X. Liu, M. Dobrowolska, J. K. Furdyna, K. M. Yu, W. Walukiewicz, G. B. Kim, M. Cheon, X. Chen, S.M. Wang, and H. Luo, *Appl. Phys. Lett.* **82**, 4310 (2003).

[16] R. P. Panguluri, G. Tsoi, B. Nadgorny, S. H. Chun, N. Samarth, and I. I. Mazin, *Phys. Rev. B* **68**, 201307 (2003).

[17] J. B. Forsyth and P. J. Brown, J. Phys.: Cond. Matt. **2**, 773 (1990).

[18] Note that the PCAR technique cannot distinguish between different signs of spin polarization.